\documentclass[aps,prl,twocolumn]{revtex4}
\usepackage{epsfig}

\begin{document}

\newcommand{\be}{\begin{eqnarray}}
\newcommand{\ee}{\end{eqnarray}}
\newcommand{\beq}{\begin{equation}}
\newcommand{\eeq}{\end{equation}}
\newcommand{\xx}{\begin{eqnarray*}}
\newcommand{\yy}{\end{eqnarray*}}
\newcommand{\nn}{\nonumber}
\newcommand{\Vol}{{\rm Vol}}
\newcommand{\sign}{{\rm sign}}
\newcommand{\tr}{{\rm Tr}}
\def\cut{{}\hfill\cr \hfill{}}

\title{Intermittency and non-Gaussian fluctuations \\
 of the global energy transfer in fully developed turbulence}

\author{ B. Portelli,  P.C.W. Holdsworth and J.-F. Pinton}

\affiliation{ Laboratoire de Physique, Ecole Normale Sup\'erieure,
46 All\'ee d'Italie, F-69364 Lyon cedex 07, France}

\begin{abstract}
We address the experimentally observed non-Gaussian fluctuations
for the energy injected into a closed turbulent flow at fixed
Reynolds number. We propose that the power fluctuations mirror the
internal kinetic energy fluctuations. Using a stochastic cascade
model, we construct the excess kinetic energy as the sum over the
energy transfers at different levels of the cascade. We find an
asymmetric distribution that strongly resembles the experimental
data. The asymmetry is an explicit consequence of intermittency
and the global measure is dominated by small scale events
correlated over the entire system. Our calculation is consistent
with the statistical analogy recently made between a confined
turbulent flow and a critical system of finite
size.

\end{abstract}

%\pacs{47.27.-i,64.60.Cn}

\maketitle

%%% INTRO %%%%%%%%%%%%%%%%%%%%%%%
%%%%%%%%%%%%%%%%%%%%%%%%%%%%%%
%%%%%%%%%%%%%%%%%%%%%%%%%%%%%%
 Intermittency is a well established feature of turbulent flows.
It has been observed in a large number experiments and
incorporated in many models~\cite{Frisch,monyag}. In this paper
we attempt to link the statistics of the fluctuations of the global
energy consumption $P_I$ in a turbulent flow to the intermittency
of the energy transfer.
Our motivation stems from the empirical observation~\cite{BHP} that the
probability density
function (PDF) of power consumption $P_I$ for a confined
turbulent flow driven at constant Reynolds number
$Re$ has a non-Gaussian form, asymmetric with an exponential
tail~\cite{LPF,PHL}, extremely similar to that observed for the order
parameter fluctuations in a
finite size equilibrium system at criticality.
Very similar distributions can be observed
for global quantities in other
experimental~\cite{Lathrop,Pumir,Car,JAN} and model correlated
systems~\cite{PRL}.  These observations have generated
recent interest in diverse aspects of global
fluctuations, for systems both in and out of
equilibrium~\cite{aji.01,fauve.01,farago,noullez,antal,dan}.
We have previously argued that the observation of similar fluctuations
 in the global quantity for these radically different correlated systems
implies an analogy at the statistical level~\cite{BHP}. Specifically, we
have
shown~\cite{PRE} using the low temperature phase of the 2D-XY model,
 that a key ingredient for
the occurrence of such non-Gaussian fluctuations is the action of
modes with a diverging distribution of spin wave stiffness across
spatial scales. In order to apply these findings to the statistics
of power injection in a confined turbulent flow, we consider
turbulence in the framework of a stochastic cascade across scales.
The flow state, at a statistical level is viewed as gas of
independent fluctuations at all scales $\ell$ between the
injection and dissipative lengths, $\eta$ and $L$.
%The energy flux
%in the flow results from the contribution of all scales.
The
variance of the fluctuations of these modes of energy transfer
diverge at small scale, which is the well accepted feature of
turbulence, called intermittency~\cite{Frisch}. The divergence
across the scales is analogous to that observed in the model
critical system.  In this paper we therefore show, using a
phenomenological model, that  the observed non-Gaussian
fluctuations of the global power injection are a natural
consequence of intermittency.

%%% Experiment and phenomonology %%%%%%%%%%%%%%%
%%%%%%%%%%%%%%%%%%%%%%%%%%%%%%%%%%%
%%%%%%%%%%%%%%%%%%%%%%%%%%%%%%%%%%%
Experimental data, reproduced from reference~\cite{PHL}, are shown
in figure 1. The flow is driven by two concentric, counter rotating discs,
with cylindrical geometry,
rotating at constant frequency $\Omega$.
The probability distribution for the fluctuations in the power, $P_I(t)$,
needed to drive the flow,
is characterized by a negative skewness,
with an approximately exponential tail for fluctuations below the mean
and a much more rapid fall-off for fluctuations above the mean.
The energy
balance is given by~\cite{LPF}:
\begin{equation}
\frac{dE(t)}{dt}=P_I(t)-P_D(t) \ ,
\end{equation}
where $E(t)$ and $P_D(t)$ are respectively the instantaneous
kinetic energy and dissipation by viscous forces~\cite{monyag,LPF}
of the confined fluid. A time average of equation~(1) yields the
condition of stationarity:
$\overline{P_I}=\overline{P_D}=M\epsilon_0$, where we define
$\epsilon_0$, the power consumption of the flow per unit mass, and
$M$,  the total mass of the fluid.

In the absence of a microscopic description of the statistical properties
of the flow, a phenomenological model is required to explain the observed
fluctuations about this mean value:
 firstly, we propose that the  fluctuations in $P_I(t)$
are related to fluctuations in the internal kinetic energy of the flow
\begin{equation}
P_I(t) = M\epsilon_0 -{\cal{E}} (t)\ .
\end{equation}
Here, ${\cal{E}} (t)= (E(t) - \overline{E})/\tau$ is the
difference between the instantaneous and mean kinetic energy,
normalized by $\tau$, an integral time scale characteristic of the
energy injection and of the response of the driving mechanism
(inertia).  That is, we propose that the torque required to shear
a fluid at constant rate $\Omega$ is a decreasing function of the
internal kinetic energy of the fluid. We predict, therefore, that
the constraint, $\Omega=const.$, forces the fluctuations in
$P_I(t)$ to be a  mirror reflection of the fluctuations of the
internal energy of the flow. If constrained with constant torque
one would consequently expect reversed power fluctuations with
positive skewness, as also observed experimentally~\cite{cadot}.
We do not suppose that mechanical structures drive the disks,
simply that the engagement of the disks is reduced with increasing
kinetic energy of the underlying flow. An excess of kinetic energy
should be repartitioned over the entire flow and felt
simultaneously at the two disks, which would then decrease their
engagement in the fluid, independently of their rotation
direction. This scenario provides a stabilizing loop through which
the turbulent steady state is maintained and it is consistent with
experimental observation that the time variation of the power
injected into the upper and lower discs displays  positive
correlations:  power fluctuations of all amplitudes are felt
quasi-simultaneously at both disks, despite their rotation in
opposite directions (See Fig. 1a of ref.~\cite{PHL}).

Secondly, we build  ${\cal{E}} (t)$ as the excess of transferred
energy at time $t$, summed over contributions from all scales in
the flow. For this we introduce $\pi_{\ell} (\vec{r},t)$, the
instantaneous real-space energy flux:
\begin{equation}
\pi_\ell (\vec{r}, t) = - \frac{1}{4} \vec{\nabla}_{\ell} \cdot
\left( | \delta_\ell \vec{u}(\vec{r}) |^2 \delta_\ell
\vec{u}(\vec{r}) \right) \ , \label{eqpil}
\end{equation}
which we average over the flow volume to obtain the energy flux at
scale $\ell$
\begin{equation}
\pi_\ell (t) = \frac{1}{V} \int_{V} d^3r \;  \pi_\ell(\vec{r}, t)
\ ,
%\label{eqpil}
\end{equation}
these variables could be computed in a direct numerical
simulation. Our phenomenology concerns the fluctuations away from
the steady state at all scales $\ell$: $(\pi_{\ell}
(t)-\epsilon_0)$. We adopt a statistical point of view rather than
a dynamical one and we make the assumption that the $(\pi_{\ell}
-\epsilon_0)$ are statistically independent stochastic
variables~\cite{Peinke} (one should think of this as a non-linear
change of variables rather than a physical tree structure
replacing the real flow). We then define the excess energy
${\cal{E}}$ as the sum over contributions from all scales:
\begin{equation}
{\cal{E}}=M\sum_{\ell} (\pi_{\ell}-\epsilon_0) \ . \label{eqE}
\end{equation}
This is a strong approximation, whose justification rests with the
results it produces, which are in striking agreement with
experiment-see Fig. 1. Physically, our model implies that the
energy injection at a given time takes into account the entire
structure of the energy flow at that time. When $\pi_\ell(t) >
\epsilon_0$, energy is built up at scale $\ell$ and the overall
energy excess ${\cal{E}}(t)$ is the sum of such effects over the
entire range of scales.

Equation (\ref{eqE}), together with our assumptions allows one to
construct the PDF of ${\cal{E}}$ as the convolution of probability
densities for the elements $(\pi_{\ell}-\epsilon_0)$:
\begin{equation}
\Pi({\cal{E}}) = \prod^{\otimes \ell} p_\ell(\pi_{\ell}-\epsilon_0) \ ,
\label{conv}
\end{equation}
where $p_{\ell}(\pi_{\ell})$ is the PDF for the fluctuations of
$\pi_{\ell}$.  Once the $p_{\ell}(\pi_{\ell})$ are chosen, the above
expression allows for the numerical computation of the
distribution for the global quantity.
In fact, one can express the
PDF for the normalized variable $\phi=\frac{\cal{E}-
<\cal{E}>}{\sigma}$ as
\begin{equation}
\Pi(\phi)=\int_{-\infty}^{+\infty}\frac{dk}{2\pi} e^{ik\phi}\Psi(k)
\label{pmu} \ ,
\end{equation}
where the characteristic function $\Psi(k)$ is the product of the
characteristic functions of each of the microscopic distributions
and the net variance $\sigma^2$ is the sum of the microscopic
variances. Then, using
 equation~(2), the PDF for the fluctuations of the
normalized power input
($\theta = \frac{P_I - \overline{P_I}}{\sigma_{P_I}}$) is
$\Pi(\theta) = \Pi(-\phi)$.
%\textsf{ A comment about why choosing $p_\ell(\pi_\ell)$ is  the
%equivalent of a Gibbs postulate for turbulence. How posh.}

%%%%%%%%%% la suite %%%%%%%%%%%%%%%
%%%%%%%%%%%%%%%%%%%%%%%%%%%%%
%%%%%%%%%%%%%%%%%%%%%%%%%%%%%
The phenomenology proposed here cannot be developed within the context of
Kolmogorov's $1941$ (K41) theory  of turbulence~\cite{Frisch}, as it
ignores fluctuations around mean values of energy injection and transfer.
We therefore develop equation~(\ref{eqE}) in the context of Kolmogorov's
Refined Similarity Hypothesis~\cite{monyag,kraichnan}.
It allows for anomalous scaling of the energy transfer at scale $\ell$:
\begin{equation}
\langle \pi_{\ell}^q \rangle \propto \epsilon_{0}^q \;
\left( {\ell}/{L} \right)^{\tau(q)} \ .
\label{eps}
\end{equation}
The spectrum $\tau(q)$ is related to the experimentally observed
non-linearity in the scaling exponents of the longitudinal
velocity increments: $\langle \delta_{\ell}u^q \rangle  \propto
{\ell}^{\zeta(q)}\propto \langle \pi_{\ell}^{q/3} \rangle
\left( {\ell}/{L} \right)^{q/3}$ so that $\zeta(q) =
q/3+\tau(q/3)$.

%%%%%%% et on continue %%%%%%%%%%%%%%
%%%%%%%%%%%%%%%%%%%%%%%%%%%%
%%%%%%%%%%%%%%%%%%%%%%%%%%%%
In the following we use two different forms for the microscopic
distributions, compatible with the experimentally determined
exponents for the velocity structure functions $\zeta(q)$. We show
that intermittency is necessary to obtain asymmetric non-Gaussian
distributions with negative skewness. However, let us first return
to some details of the model: the $(L/\eta)^3$ correlated degrees
of freedom implicated in the flow~\cite{monyag} are transformed
into an ensemble of $N$ scales. At each level the system is
described with a resolution length scale $\ell_{n}$ and the scales
are separated by a ratio $\lambda = (\ell_{n-1}/\ell_n)^3$ such
that $\lambda^{N}=(L/\eta)^3$. The Reynolds number and the number
of levels are related through the definition $Re = (L/\eta)^{4/3}
= \lambda^{4N/9}$. Note that the variable $\lambda$ is a free
parameter in the theory. In the original Kolmogorov-Obukhov theory
(KO62) of intermittency~\cite{monyag}, the energy cascade, being a
multiplicative process with a large number of steps, is assumed to
have log-normal statistics. In this case $p_n(\log \pi_{\ell_n})$
is a Gaussian distribution whose variance can be that of KO62 or
that proposed by Castaing {\it et al.} \cite{cast}. As one assumes
scale invariance in KO62, the variance is $\sigma^2_{n} =
\epsilon_{0}^2 \left[ ({\ell_n}/{L})^{\tau(2)}-1\right]$. This
distribution leads to the quadratic spectrum $\zeta(q)= q/3 + \mu
q(3-q)/18$.  The value of  the intermittency parameter $\mu \equiv
-\tau(2)$ that  fits best the experimental data is $\mu =
0.21$~\cite{arneodo}. Another possibility, which gives a very good
fit to the velocity intermittency exponents  is a $\chi^2$
distribution:
\begin{equation}
p_{n}(\pi_{\ell_{n}})=N_{\ell_n}\pi_{\ell_{n}}^{\nu_{\ell_{n}}-1}\exp{(-a_{\
ell_{n}}\pi_{\ell_{n}})}.
\label{pchi2}
\end{equation}
In this case, we require that the
variance of $p_{n}(\pi_{\ell_{n}})$ be that of KO62,
in addition to the conditions of normalization and constant mean. The
three conditions determine $N_{\ell_n}$, $a_{\ell_n}$,
$\nu_{\ell_n}$. From the expression for the moments
($
<\pi_{\ell_{n}}^q> = \epsilon_{0}^q   \Gamma(q+\nu_{\ell_{n}})  /
\nu_{\ell_{n}}^q  \Gamma(\nu_{\ell_{n}})
$)
it is straightforward to show that the spectrum $\tau(q)$ is that
of KO62 in the limit $\ell_{n}\longrightarrow L$. Even if $\chi^2$
statistics does not strictly  allow for scale invariance for
$\tau(q)$ and $\zeta(q)$ in the inertial range, the corrections
are very small. The main advantage of
the $\chi^2$ distribution is that it allows a
straightforward calculation of the convolution product in
equation~(\ref{pmu}):
\begin{equation}
\log\Psi(k)=ik\epsilon_{0}/\sigma-\sum_{n=1}^{N}\nu_{\ell_{n}}
\log(1+ik/(\sigma a_{\ell_n}))
\end{equation}

%%% INTERMITTENCY : PDF SHAPE, VARIATION WITH RE %%%
\begin{figure}[ht]
\vspace{-0.5cm}
\epsfig{file=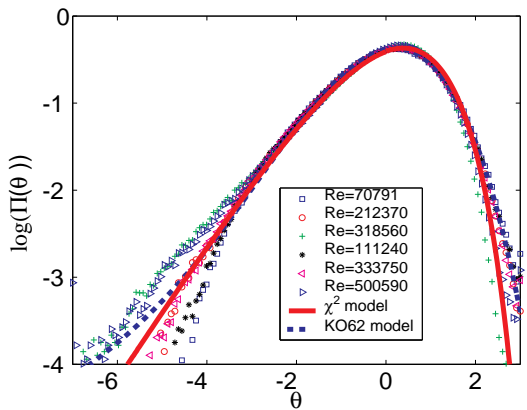,width=7cm,height=5.5cm}
%\vspace*{45mm}
\caption{PDF of
global energy transfer for experimental data, taken from Pinton
{\it et al.}~\cite{PHL} (symbols) and from cascade models of
intermittency. The cascade models are: $\chi^2$-model, $Re=10^5$
and $\lambda=2.0$ (solid line) and lognormal model, $Re=10^5$ and
$\lambda = 1.08$ (dashed line).} \label{pdf-1}
\end{figure}
%\vspace{-0.3cm}
In figure 1 we show the experimental data published in
ref~\cite{PHL} together with the PDF  for $\Pi(\theta)$, calculated from
equation~(5), using both
log-normal and $\chi^2$ models. They both capture the essential
features of the experimental data: (i) the fluctuations in $\theta$
are strongly non-Gaussian, despite the high value of the Reynolds
number, i.e. of the large number of contributions in the
sum~(\ref{eqE}); (ii) the distribution is skewed towards negative
values; (iii) the tails of the PDF towards
negative values are almost  exponential, although this seems
better verified for the $\chi^2$ microscopic distribution than
for the log-normal. This behavior is extremely robust against variations of
the
Reynolds number, at fixed $\lambda$. In this case, increasing the
Reynolds number leads to an increase of the number of cascade
steps $N$ (as $\log Re$) and, at the same time, to an increase of
the asymmetry in the shape of the microscopic distribution at the
smallest scales. In figure~2 we show the evolution of the
distribution with $Re$, for $\lambda = 2.0$, for the $\chi^2$
microscopic distributions. The curves vary only slowly with $Re$.
The asymmetry becomes more pronounced as $Re$ increases, as can be
seen from the skewness, $\gamma$, which varies from $-0.64$ to
$-0.97$ for $Re$ varying from $10^4$ to $10^7$. Hence,  as in the
experiment, there is no evidence that $\Pi(\theta)$ will reduce to a
Gaussian in the limit of infinite Reynolds numbers, and therefore
infinite $N$. Given the
uncertainty in the experimental results and the crudeness of our
model, this qualitative agreement seems extremely
encouraging.

As $\lambda$ is a free parameter in the problem, it has been
chosen in figure~1 to give a good fit of the experimental
curves for each class of microscopic distribution. For a fixed
value of $Re$, varying $\lambda$ does not change the shape of the
global PDF. Using the $\chi^2$ distribution, we observe a slight increase of
the skewness, from
$-0.65$ to $-0.83$ when $\lambda$ is increased from $1.5$ to $2$ (for
Re=$10^6$), a
typical range of values used in cascade models~\cite{Frisch}.

\begin{figure}[ht]
%\vspace{1cm}
%\epsfig{file=fig2-PPRL.eps,width=7cm,height=5.5cm}
\epsfig{file=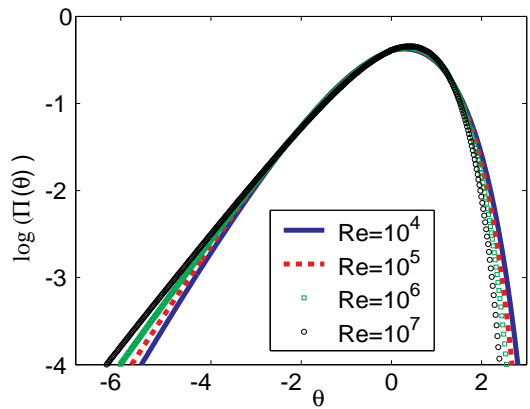,width=7cm,height=5.5cm}
%\vspace*{45mm}
%\vspace{-0.2cm}
\caption{PDF of global transferred energy for $Re$ varying between
$10^4$ and $10^7$ (legend). The microscopic distributions are
$\chi^2$ ones and $\lambda=2.0$. } \label{pdf-2}
\end{figure}
%\vspace{-0.5cm}
However, there seems no {\it a priori} reason to fix $\lambda$ and
while  we do not rule out a weak variation, we can fix
a universal form for the PDF, as proposed in~\cite{BHP}
by letting $\lambda$ vary
with $Re$. Support
for this proposition comes from the evolution of the ratio
$\sigma/\overline{P_I}$ with $Re$. Experimentally, one observes a very
slow decrease that may be fitted by a power law dependence with
exponent $\alpha \sim 0.3$~\cite{PHL}. In our computations at
fixed $\lambda$, we find the ratio to be slightly increasing, in
disagreement with experiment. If, instead of fixing $\lambda$ we
impose universality, we find a decreasing ratio, with a weak power
law dependence with exponent $\alpha \sim 0.1$.

%%% DISCUSSION / TURBULENCE %%%
%The quantity $\cal{E}$ involves the sum over $N$ statistically
%independent objects which would
%result in a Gaussian distribution  for $\Pi(\theta)$ if the individual
%elements were of similar amplitude. However, this is not the case, as the
%intermittency leads to a range of variances that diverges in the
%limit of infinite $Re$. This gives the possibility for violation
%of the central limit theorem and a non-Gaussian distribution for
%the global quantity defined above.
Here, the small scales  play a major role due to the dispersion in
the amplitudes  and as in the finite size critical
system~\cite{modes}, one might expect the global PDF to retain
some characteristics of the microscopic measure, with the
experimentally observed asymmetry reflecting the form of
$p_n(\pi_{\ell_{n}})$. This is indeed what we observe and the
symmetry of the global PDF is only restored if the individual
elements are made to have equivalent weight. This is illustrated
in figure~3, where we show a series of PDFs produced at fixed $Re$
and $\lambda$, with $\tau(2)$ varying between the experimental
value, $0.21$ and zero. As $\tau(2)$ is reduced the PDF becomes
more symmetric and, for $\tau(2) = 0$ it  is well represented by a
Gaussian. This point is quite important: intermittency is
absolutely necessary to obtain an asymmetric and non-Gaussian PDF
for the global energy transfer. However this conclusion only holds
for the PDF of energy transfer. In the case of pressure
fluctuations Holzer {\it et al.} \cite{holzer} have shown that
intermittency is irrelevant since for the pressure one integrates
$\delta_{l} u^2$ which gives a dispersion of variance $<\delta_{l}
u^2>\sim  (\epsilon_{0} l)^{2/3}$ in the strict $K41$ description.
We remark that, with the phenomenology proposed here, negative
skewness is a direct consequence of kinetic energy fluctuations
with positive skewness, as are observed in microscopic models with
dissipation~\cite{fauve.01,farago,noullez}.

The importance of the small scale can be gauged by recalculating
$\cal{E}$, excluding levels from either the top or the bottom of
the cascade. We find that with removal of  cascade levels near the
dissipative scale the distribution rapidly approaches a  Gaussian form,
 while removal of levels from
the top of the cascade leaves the PDF essentially
unchanged. The entire global
measure is therefore seen to be influenced by a relatively small
number of statistically independent cascade levels describing the
system at small scales. If a universal PDF for the global measure is
indeed an experimental reality, then the dispersion in amplitudes
must diverge with Reynolds number, in such a way that the
universal distribution is a ``limit'' function, valid even as $Re
\rightarrow \infty$. These arguments lead us to a
striking physical conclusion concerning the nature of a turbulent
flow: given the huge number of small scale objects  present in the flow
($(L/\eta)^3 = Re^{9/4} \sim 10^9$ for $Re = 10^4$); if the
PDF for the global measure is non-Gaussian, it can only mean that
these objects are strongly correlated in space and time. This
point seems consistent with recent measurements of long time
temporal correlations in the dynamics of  Lagrangian tracer
particles in a fully turbulent flow~\cite{Mordant}.

We finally address the point that the PDF of the injected power has a
Gaussian shape for open (non confined) flows~\cite{LPF}. The
distribution of the global quantity thus depends on the overall
size of the flow, although it is known that small  scale
intermittency characteristics do not; for instance, the scaling
exponents of the longitudinal velocity increments are identical in
open or confined flows~\cite{arneodo}.
%, if measured away from the boundary layer
We propose that in non-confined flows
several uncorrelated cascades occur simultaneously. In this case,
the probability for the overall energy transfer is given as the
convolution of $\cal{N}$ functional forms of the type $\Pi(\theta)$
computed above. This tends rapidly to a Gaussian.
Experimentally, this could be tested  by
changing continuously the ratio of the disk to cylinder radius in
a 'washing-machine' experimental setup.

%\noindent{\bf Acknowledgements}

%
\begin{figure}[ht]
%\vspace{1cm}
\epsfig{file=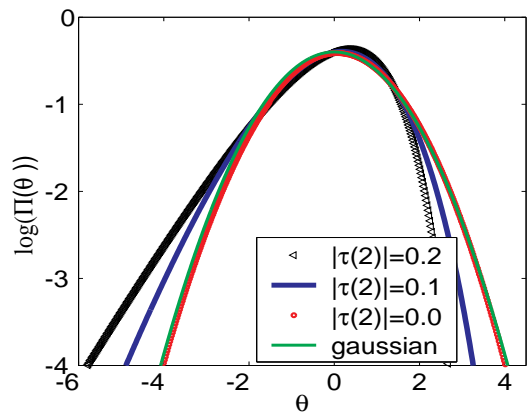,width=7cm,height=5.5cm}
%\epsfig{file=fig3-PPRL.eps,width=7cm,height=5.5cm}
%\vspace*{45mm}
\caption{Variation of the global PDF with the magnitude of the intermittency
parameter $|\tau(2)|$ for $Re=10^6$ and
$\lambda=2.0$. The microscopic distributions are $\chi^2$ statistics. }
\label{cros}
\end{figure}

The experimental data shown in figure~1, published in reference~\cite{PHL},
have been obtained
with R. Labb\'e.
We gratefully acknowledge many  fruitful discussions with M. Bourgoin,
S.T. Bramwell, B. Castaing, F. Chill\`a, T. Dombre, J. Farago, Y. Gagne,
 E. L\'ev\`eque, P. Marcq, A. Naert,  J. Peinke.
This work has been supported by CNRS ACI grant no. 2226.

%\vspace*{-0.5cm}

\end{document}